# Methods of quantifying specialized knowledge and network rewiring

Sirui Wang[*,1], Michael Macy[2,3], Victor Nee[3]

ABSTRACT

Technological innovations are a major driver of economic development that depend on the exchange of knowledge and ideas among those with unique but complementary specialized knowledge and knowhow. However, measurement of specialized knowledge embedded in technologists, scientists and entrepreneurs in the knowledge economy presents an empirical challenge as both the exchange of knowledge and knowledge itself remain difficult to observe. We develop novel measures of specialized knowledge using a unique dataset of longitudinal records of participation at technology-focused meetup events in two regional knowledge economics. Our measures of specialized knowledge can be further used to quantify the extend of knowledge spillover and network rewiring and uncover underlying social mechanisms that contribute to the development of increasingly complex and differentiated networks in maturing knowledge economies. We apply these methods in the context of the rapid morphogenesis of emerging regional technology economies in New York City and Los Angeles.

[*] Corresponding author
[1] Department of Operations, Information and Decisions, The Wharton School, University of Pennsylvania, Philadelphia, Pennsylvania, United States of America
[2] Department of Information Science, Cornell University, Ithaca, New York, United States of America
[3] Department of Sociology, Cornell University, Ithaca, New York, United States of America

*Introduction*

Entrepreneurs are often strategic in the exchange of ideas as they search for novelty, and innovation often comes from unique applications or combinations of useful knowledge embodied by individual technologists in specialized clusters of the knowledge economy. However, measurement of useful knowledge embodied in networks of technologists involves a novel methodological approach. Moreover, such measurement must take into account how individual technologists and entrepreneurs engage in selective repeated exchange with one another to facilitate the spillover of useful knowledge and knowhow in an emerging regional knowledge economy. In this article, we study groups of entrepreneurs and technologists who engage in exchange of useful knowledge and knowhow through regular face-to-face meetup events in two regional metropolitan economies. Using an online longitudinal data set of event attendance data at technology-focused events from the professional networking platform, Meetup.com, we develop a set of novel methodologies to collect behavioral traces and measurement of the effects of knowledge spillover and network rewiring as social processes that enable and motivate cooperation, trust and prosocial behavior facilitating innovative activity leading to new combination and recombination of useful knowledge. Our data allows us to measure topics of specialized knowledge for technologists and entrepreneurs who regularly participate in knowledge spillover events. We further track the influx of new actors with new knowledge and how exploration in search for useful knowledge changes patterns of social interactions over time as technologists and entrepreneurs "rewire" their networks. This paper lays out the novel methodologies we used in the second part of our explanation sketch in "A Theory of Emergence" where we focus on the social dynamics of emergence of the knowledge economy. The present article proceeds as follows: we first provide a sketch of Meetup.com, from which online data for this paper was collected; we next describe how we use these data to measure specialized knowledge and rewiring; we conclude with a discussion of the implications of our methods and findings.

*Meetup.com context and networks view of technologists in a regional tech economy*

We use data from Meetup.com to construct social networks that map the emergence and growth of regional knowledge economies in New York and in Los Angeles. Meetup.com is a New York-based professional networking platform that was founded in 2002 and has grown to reach over 49 million users worldwide.[4] Since its founding, Meetup has served as a locus for entrepreneurs of startup tech firms to network with technologists in specialized knowledge clusters. As a core economic institution of the 21$^{st}$ century knowledge economy, Meetup.com has been instrumental in facilitating face-to-face knowledge spillover events dedicated to showcasing and exchanging useful knowledge (DellaPosta and Nee 2020). Individuals create their own profiles on the platform and can affiliate with numerous special interests by either listing their interests, joining specialized groups, or attending events. Meetup's online data offers a novel opportunity to measure various types of specialized knowledge and track dynamic interactions among technologists who attend Meetup events during the formative period of emergence and rapid growth of two regional knowledge economies. Figure 1 shows an example profile on Meetup.com and potential fields that signal an individual's knowledge specialization.

We focus on the metropolitan areas of New York and Los Angeles to detail methods used in the measurement of social interactions of technologists as they venture beyond their immediate networks to explore Meetup events in search for useful knowledge. New York is the first city in which Meetup.com operated and it continues to maintain a strong presence in the regional technology community. Los Angeles is the second largest metropolitan area in the United States and provides a contrast to the urban geography of New York, spanning a much larger area connected with freeways that may not be conducive to frequent face-to-face meetings.

---

[4] Newcomb, Alyssa. "Meetup was a darling of the tech industry. But can it survive WeWork?" Dec. 26 2019. *NBC News.* https://www.nbcnews.com/tech/tech-news/meetup-was-darling-tech-industry-can-it-survive-wework-n1106676

**Figure 1:** Measuring specialization using group memberships and public profiles on Meetup.com. Circled in red are the self-listed interests and group memberships that signal an individual's knowledge specialization

(a) Public profiles containing self-reported interests

[Screenshot of a Meetup profile page for "John Smith", located in New York, NY, Meetup member since April 14, 2014. Member of 3 Meetups: The New York Python Meetup Group, NY Tech Meetup, NYC Startup Community. Interests (circled in red): Professional Networking · Programming in R · Big Data · Data Mining · Applied Statistics.]

(b) Topics associated with each Meetup group that a member joins may can the interests of that member

[Screenshot of the NY Tech Meetup group page: Brooklyn, NY; 61,728 members · Public group; Organized by NY Tech Alliance and 2 others. Related topics (circled in red): New Technology, Web Technology, Futurology, Entrepreneurship, PDF.]

For New York, we first compiled a list of all Meetup groups based in either Manhattan or Brooklyn. For Los Angeles, we compiled a similar list of all Meetup groups labeled 'tech' within a 20 mile radius of Los Angeles (including Santa Monica, Pasadena, Culver City, etc.) For each of the 'tech' Meetup groups, we collected a list of all individuals who had RSVP'ed "yes" to at least one of the events hosted by a 'tech' group. The pairwise correspondence between individuals and events generated a set of dynamic networks where nodes represent individual members and edges represent overlap in event attendance between members.

Figure 2 shows yearly snapshots of the networks of technologists in both New York and Los Angeles.[5] The figure shows rapid growth as both regional networks become increasingly dense. There is also increasing turnover (measured year-to-year) as new technologists participate in knowledge events and previous ones exit.

---

[5] Edges in these networks are weighted to distinguish between stronger and weaker connections. An edge is weighted by the cosine similarity between the two individuals from the TF-IDF (term frequency-inverse document frequency) weightings of the incidence matrix for each network. The combination of TF-IDF and cosine similarity to construct two-mode networks have also been previously explored in social science research (e.g., Hoberg and Phillips 2016, Hoffman 2019).

**Figure 2:** Visualization of the evolution of the networks of technologists in New York and in Los Angeles (200-2019)

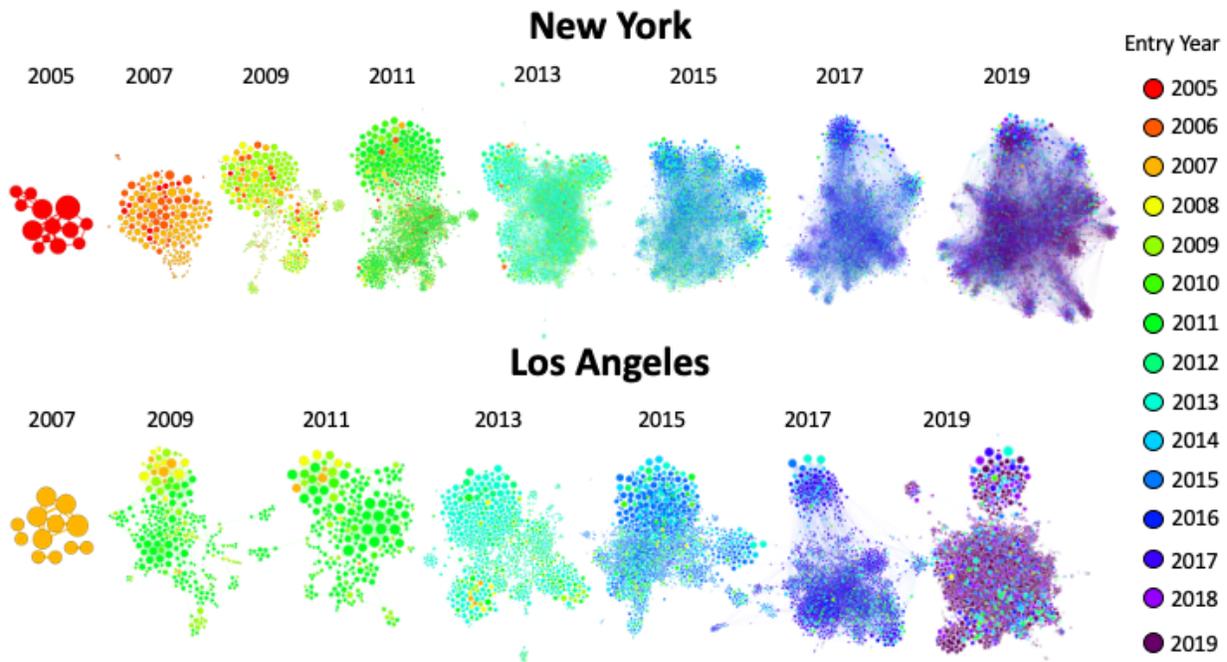

*Measuring specialized knowledge through affiliated interest terms*

An opportunity for knowledge spillover occurs when individuals with different knowledge profiles interact through one of the Meetup events they mutually attend. Although individuals with similar profiles cooperate in knowledge sharing, the discovery of a novel combination or recombination is more likely when individuals with different knowledge profiles experience knowledge spillover.

Specialized knowledge is indicated by the clustering of a user's topical interests. Users on Meetup.com can express their interests by directly listing relevant keywords on their member profile (illustrated in the sample member profile shown in Figure 1. These lists enable users to more easily connect with others who share similar interests. In Meetup "tech" events, profiles often list technology-related topics such as software development, entrepreneurship, AI or blockchain.

Because interests are self-reported and listing is not required, these data can be missing for some individuals. We supplemented the self-reported data with topics related to an individual's memberships in groups. Meetup groups play a central role in facilitating connections among members. Groups are usually organized around a common theme, identified by "topic tags" that attract participation by users who share that interest. Individuals also join groups in order to attend events that are only available to members. For each individual, self-reported interests and topic tags associated with Meetup groups comprise the *interest terms* for that individual.

The change in colors across the networks in Figure 2 suggest high rates of turnover. Despite individuals entering and exiting knowledge clusters within a short period of time, the cumulative effect may still result a net influx of new knowledge that persists among those members who remain in the network. In fact, we observe this phenomenon by measuring the average tenure of members who are associated with each interest year over time. A net influx of knowledge that persists with whoever remains active in the overall knowledge network would correspond to an increasing trend in the average tenure of members who are associated with each interest term in the network above and beyond what is expected simply due to the passing of time. We measure tenure of a member as the number of years since a member's first technology Meetup event. For example, a member whose first tech Meetup event was in 2007 will have a tenure of eleven years in 2018. In 2006, 1,712 new interest terms were introduced into the New York knowledge network; 1,678 (98.0%) of these interest terms appeared exclusively in profiles of people with zero years tenure (people who entered the network in 2006). Every interest term in each year can be characterized in terms of the average tenure of its adopters. Figure 3 is a scatterplot of average tenures for every interest term each year. Each point in the plot represents an interest term in a corresponding year; interest terms that just entered the network in the given year are represented by blue circles, while pre-existing interest terms are represented by orange triangles. The trend lines show the average of all the data points of the given type in each year, and the solid black line shows the average tenure of all members in the network in that year. Figure 3 shows that adopters of newly introduced interest terms are generally much newer to the network compared to adopters of existing interest terms as well as to the network-wide average. This indicates that new interest terms are predominantly appearing in the profiles of new members, or members with lower tenure, supporting the idea that the flow of new knowledge is associated with newer members in the community rather than exploration of new ideas by established members.

**Figure 3:** Average tenures of interest adopters based on whether the term is a new entrant into the network. Newly introduced interests are adopted by newer members compared with the network average, and older interests are adopted by more senior members compared with the network average.

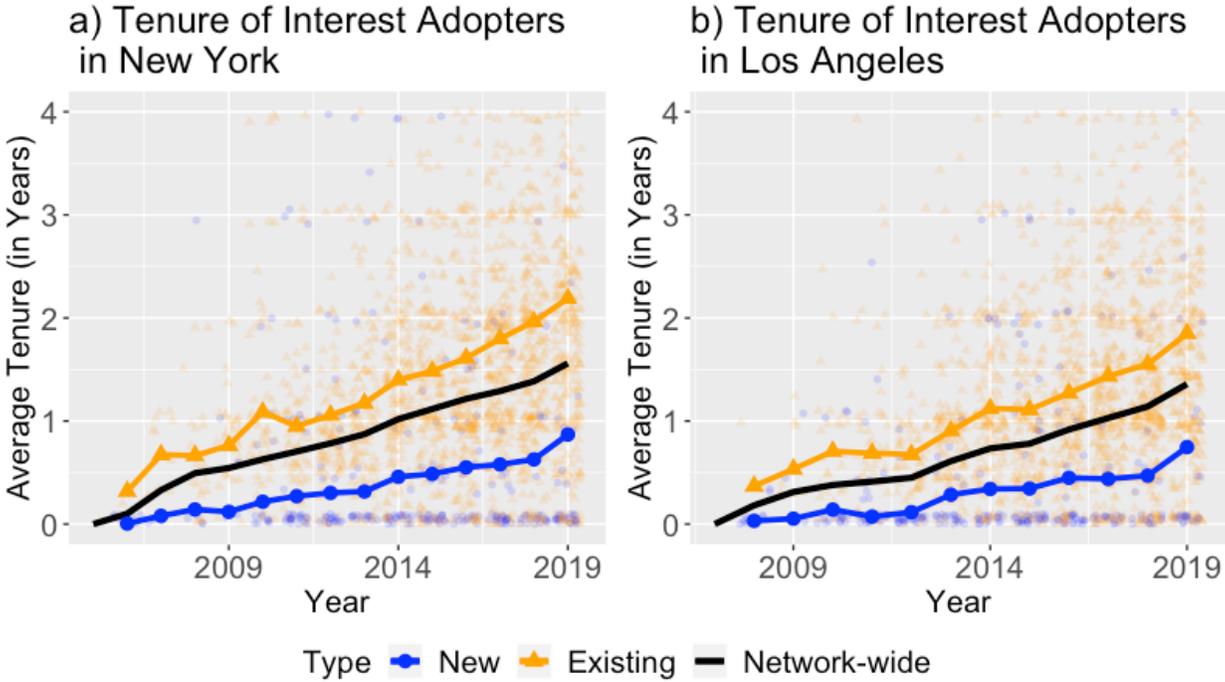

*Rewiring is associated with less specialized local networks*

Figures 2 and 3 suggest that the technologist networks in both New York and Los Angeles significantly change over time as new individuals (and knowledge) enter the network and existing individuals change the way they interact with others in the community. We develop a measure of "novelty" at the individual level to capture how a person's event attendance behavior differs from one year to the next. An individual may become less active in certain types of tech events but become more involved in other types of events over time.

Novelty corresponds to the concept of "network rewiring." Rewiring occurs when individuals explore knowledge clusters beyond their immediate network neighbors. The more novel one's exposure to the knowledge network in a given year, the more likely that their network has "rewired" from the past year.

We measure an individual's level of rewiring from year to year by comparing the groups that hosts the events the individual attended one year with those that hosted the events attended the previous year. Formally, we can represent every individual's group event attendance in year $t$ as a vector, $\boldsymbol{x}_t$, of size $K$ where $K$ is the total number of technology Meetup groups. Each element $x_t^{(k)}$ corresponds to the count of events that the individual has attended in year $t$ hosted by Meetup group $k$. For example, if the individual attends 5 events hosted by the "Blockchain Meetup" and 2 events by the "Self-driving Vehicles Meetup" in 2017, and the same individual attends 7 events hosted by the "Blockchain Meetup" and 1 event hosted by the "Deep Learning Meetup", we can represent this attendance record as in Equation 1.

$$\begin{array}{cccc} & Blockchain & SelfDriving & DeepLearning \\ \boldsymbol{x}_{2017} = ( & 5 & 2 & 0 \quad) \\ \boldsymbol{x}_{2018} = ( & 7 & 0 & 1 \quad) \end{array} \quad (1)$$

We calculate the "event novelty score" of an individual in year $t$ as 1 minus the cosine distance between the the individual's attendance vector in year $t$ and that in year $t-1$. This is formally stated in Equation 2.

$$Novelty_t = 1 - \frac{x_t^T x_{t-1}}{\|x_t\|\|x_{t-1}\|} \quad (2)$$

An individual's novelty score is bounded between 0 and 1; if a technologist attends the same events hosted by the same groups in the same proportions each year, their novelty score will be consistently 0. On the other hand, if a technologist only attends events hosted by groups with which they have never previously interacted, their novelty score will be consistently 1. Any observed novelty scores for individuals fall somewhere between these two extremes. Figure 4a shows the distribution of observed novelty scores for individuals over time in the two regional networks. On average, we observe increasing levels of event novelty among technologists in both metropolitan areas while the total number of events attended by each individual appears to hold constant. This suggests that individuals maintain a stable level of engagement with Meetup events, but the events they attend change over time such that individuals continually have the opportunity to explore new knowledge.

**Figure 4:** Distribution of event novelty scores over time for technologists in New York and Los Angeles. The aggregate level of event novelty increases over time, though the number of events attended by each person holds steady, suggesting that the events they attend are changing from year to year as they explore outside their existing relationships.

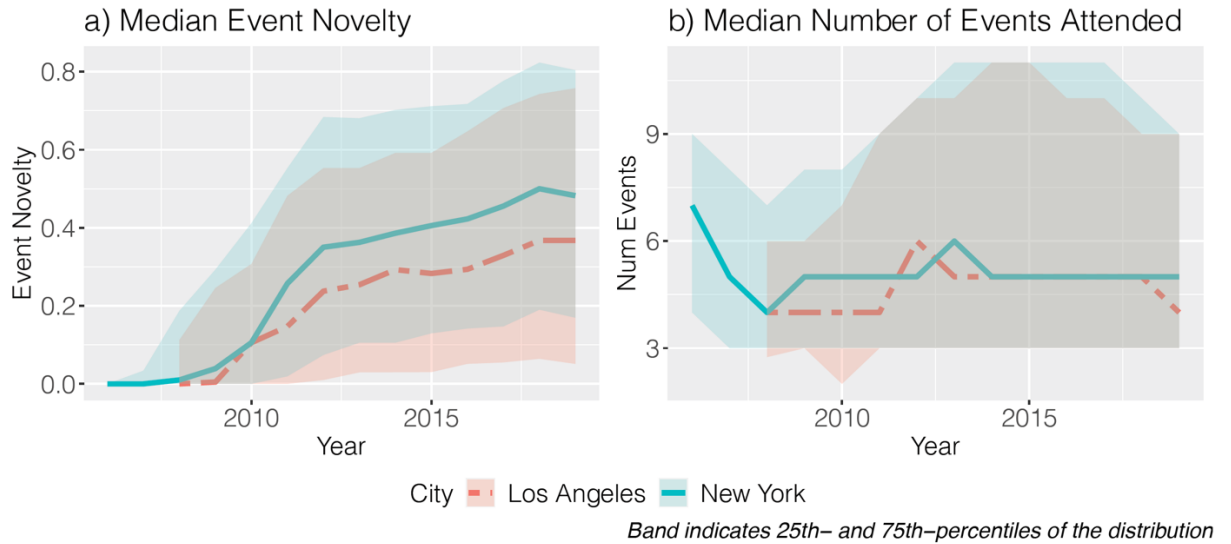

Entrepreneurial actors in search of novelty rewire their networks as they access a broader range of technologists. As a result, social interactions among technologists attending the same or similar knowledge spillover events motivate rewiring of networks, more so than when players stay in place in a static network. Diversity of interests of network neighbors can be measured by how much the distribution of interests of an individual's neighbors deviates from that of the entire network. The more an individual changes his or her event attendances, the more the individual will be exposed to less specialized interests that are more representative of the broader population.

"Specialization" refers to the degree to which a set of interests are skewed towards a particular focus relative to the overall distribution of knowledge in the network. Specialized local networks reflect parts of the network that are focused on a limited set of ideas. Individuals who rewire more extensively are also more likely to be positioned in less specialized local networks.

An individual's level of specialization relative to the broader population is quantified as the distance between the interest distributions of the individual's ego network and the entire population. More precisely, an individual's specialization score is the sum of the absolute difference between the proportion of each interest as observed in an individual's ego network and the proportion of the interest in the overall population. (A formal description of the calculation of the specialization score is given in Appendix A.) The higher the specialization score, the more the interests of a member's social circle differ from that of the general population.

This measure is context specific. We observe the specialization profiles of every individual and those of his or her network neighbors and compare that profile to the aggregate profile of the entire population. For example, if an individual's specialization profile is mostly characterized by terms related to AI, and he or she also attends events with others who are predominantly interested in AI, then the profile of that individual's ego network will be heavily skewed towards AI interest terms when compared with the entire population, where AI may be just one of a number of equally popular terms.

Using this ego-network specification, we can track a member's degree of specialization through the technologists with whom he or she connects over time as the regional technology economies evolve. The introduction of new interest terms into the network over time can affect specialization scores in both directions. Rewiring as a mechanism to access diverse knowledge occurs when an individual bridges across different specialties, such that their ego-network comprises an expanded set of unique interests. This in turn deflates the individual's specialization score. Equation 5 models the effect of year and rewiring on individual's specialization score, controlling for individual fixed effects, number of events attended, and total number of event co-attendees (connections) that year.

$$\text{SpecializationScore}_{it} = \beta_0 + \beta_1 \text{Year}_t + \beta_2 \text{EventNovelty}_{it} + \boldsymbol{\beta}\text{Controls}_{it} + \alpha_i + \epsilon_{it} \quad (2)$$

**Table 1:** Coefficient estimates of a panel regression model for ego-network specialization scores over time on individual-level characteristics. Significance is calculated using heteroskedasticity

robust standard errors. For both New York and Los Angeles, more rewiring is associated with a lesser degree of specialization among acquaintances in an individual's ego-network.

|  | *Dependent variable:* | |
|---|---|---|
|  | Specialization Score | |
|  | (1) New York | (2) Los Angeles |
| Year | 0.030*** | 0.016*** |
|  | (0.001) | (0.003) |
| Event Novelty | -0.021*** | -0.025** |
|  | (0.002) | (0.004) |
| Num Events Attended (log) | 0.120*** | 0.050*** |
|  | (0.005) | (0.008) |
| Num Network Connections (log) | -0.648*** | -0.600*** |
|  | (0.006) | (0.009) |
| Observations | 88,724[1] | 25,012[1] |

\* p<0.05; \*\* p<0.01, \*\*\* p<0.001

[1]Some observations are omitted based on privacy settings of certain members that do not publicly list interest or group memberships, which are needed to calculate specialization scores

The coefficient estimates, shown in Table 1, indicate that individual specialization scores tend to increase over time. In both New York and Los Angeles, there is a significant negative covariance between rewiring and specialization scores amid a tendency for increasing specialization over time. A one standard deviation increase in event novelty is associated with a 0.02 standard deviation decrease in specialization in New York and a 0.03 standard deviation decrease in Los Angeles. These results suggest that rewiring and general exploration of the tech network allows actors to bridge across loci of specialized knowledge despite the network-wide tendency for increasing specialization.

***Rewiring accounts for more integrated structure in the knowledge network***

We use network modularity at the knowledge cluster level to measure the aggregate level of interconnectedness between and within localized clusters of individuals. In biology, modularity refers to the construction of a cell organism by joining together discrete individual units to form larger and more efficient configurations. This facilitates evolutionary change because entire modules can be rewired while maintaining their modular functions (Kirschner and Gerhart 2005). Analogously, we can study the emergence of a knowledge-based industrial cluster formation with meso-level analysis of the organizational field (Nee 2005: 56, 60-62), in particular, the rewiring activities between distinct players responding to new environmental states and opportunities (Luscombe et al. 2004). Within the knowledge economy, modularity enables innovative activity if actors in a knowledge cluster can bridge across multiple clusters of specialists who are progressively and dynamically differentiated. DellaPosta and Nee's (2020) longitudinal study of email threads of technologists in the emergence of New York City's knowledge economy shows that social processes giving rise to specialization and diversification are complementary. Specialization at the individual and group levels does not lead to balkanization of specialized knowledge communities sealed off from one another; rather, it provides a foundation for integrating diverse knowledge and knowhow.

Individual access to diverse knowledge presents an opportunity for innovative activity. We measure *opportunity* for innovative activity at the network level as modularity in the connectivity across the clusters of the knowledge network. We use network modularity to quantify the community structure of the networks over time. Intuitively, network modularity is higher for networks with more clearly defined community clusters, with a large number of edges between nodes of the same community and a small number of edges between nodes of different communities. Thus, high network modularity implies that interactions among individuals do not venture beyond local clusters, while low modularity suggests higher interconnectedness among clusters. Well-defined interconnected subcommunities in the network imply that groups of individuals are attending the same events with one another and not with others outside the group. In contrast, in networks with low modularity, every part of the network is more integrated with the rest and interactions occur more uniformly. The community structure of a network is thus sensitive to individual-level rewiring of event attendances. Comparing modularities across multiple networks allows us to quantify the differences in complexity of networks in terms of their community structure. (The formal calculation of network modularity is given in Appendix B.)

*Entry with undifferentiated preferences*

Simulations of counterfactual event attendance can isolate the effects of rewiring on the evolution of overlapping and cross-cutting links between knowledge clusters in the observed network structure in New York and Los Angeles. The simulations change the way individuals attend events and thus the way individuals connect with one another through events. Figures 2 and Figure 3 together suggest that a major source of new knowledge within the tech communities over time is the entry of new members into the network. If the influx of new members alone, rather than their preferences, is the main driver of structural change of the network of an emerging knowledge economy, we would expect to see a similar pattern of modularity if we simulated random event attendances under similar levels of new entry to the network, compared to what we actually observe to be the case in New York and Los Angeles.

Figure 5 compares the observed modularity of each network with simulated modularities of randomly generated networks that correspond to the original network. This allows us to attribute any change in modularity to the influx of new members alone. Since members differentiate themselves through event attendances, a scenario in which a member attends any event with equal probability would result in an undifferentiated network. We use this intuition as the basis of the randomly generated networks. For each year, we randomly assign each member's event attendances while preserving the total number of events that each member attends. We build a new network using this generated event attendance and calculate the modularity of the undifferentiated network using the Louvain algorithm (Blondel et. al 2008). Comparing the observed modularity with the expected undifferentiated modularity teases out the effect of preference-undifferentiated entry, since the simulated network preserves the rate of network growth as well as the distribution of the number of events each member attends.

Figure 5 shows a) the observed modularities of each network, b) the expected modularities of a comparable undifferentiated network, and c) the difference between the observed and expected modularities. Overall, we find that explorations of individual actors across Meetup groups in our sample are characterized by increasing *modularity* over time, but introducing rewiring maintains higher between-group connectivity throughout.

**Figure 5:** Observed network modularity and expected network modularity under entry with undifferentiated preferences over time for New York and Los Angeles tech communities. Community structure appears to increase over time despite the expected tendency to decrease with simple undifferentiated entry.

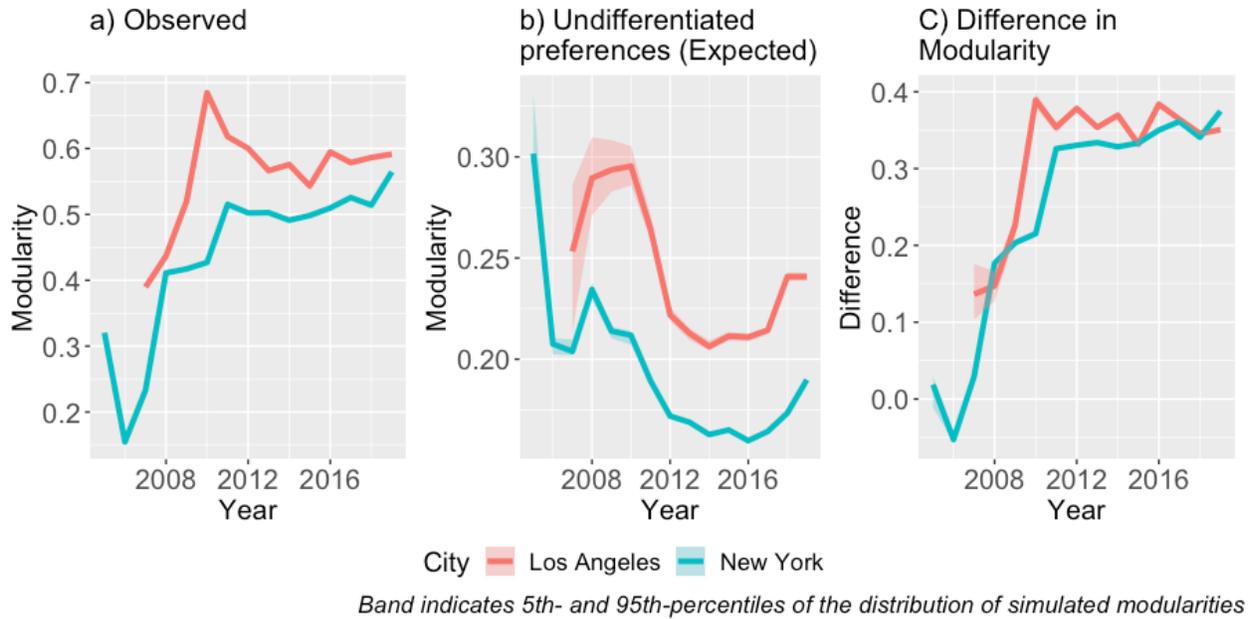

Figure 5a shows that modularity of the observed member networks has generally increased in the early years and has held steady in more recent years. Figure 5b shows that the trend for expected modularity given undifferentiated preferences would be generally decreasing, the opposite of what is observed in 5a. This suggests that in the absence of heterogeneity of preferences across members, entry of new members alone cannot explain the increasing modularity in community structure that is observed in both the New York and Los Angeles knowledge economies.

*Entry with differentiated static preferences*

Simulations of counterfactual event attendance can also be used to explore the implications of heterogeneity in attendance behavior between members, while keeping event preferences static over time (i.e., no rewiring). Any residual differences between the counterfactuals and the observed outcomes can then be attributed to rewiring. To that end, Figure 6 shows observed and expected modularities under a simulation of heterogenous attendance propensities among members that are fixed through time. Simply put, holding an individual's propensity to attend certain events constant removes within-individual variation, leaving only between-individual heterogeneity of preferences. For the yearly simulated network with no rewiring, we estimated the propensities for attending any type of event during the first year of a member's entry into the network and used that propensity to simulate event attendances for the remaining years that the individual remains in the network.[6] Members that stay in the network for more than one year do not rewire or change their propensity to attend events from year to year.[7] This is analogous to the event novelty measure of rewiring in the previous section, as individuals will, on average, have zero event novelty scores if their propensity to attend events hosted by any group remains constant across years.

      Figure 6a shows the observed modularity in the member networks, while Figure 6b shows the expected trends in modularity if the community exhibits event preferences that are heterogeneous across individuals but static within individuals. Note that this trend closely tracks the observed trends, suggesting that heterogeneous preference plays a major role in driving the changes in community structure in these networks. Figure 6c shows the differences in the observed and expected modularity trends under static differentiated preferences. We attribute this difference to rewiring, as indicated by changing event attendance behavior over time. We see that in Los Angeles and in the later years of New York, the lack of rewiring results in higher modularity compared to the observed network where rewiring is possible (the difference as shown in Figure 6c is largely negative). Unfortunately, the available data cannot distinguish if an entrepreneur or technologist changes event attendance behavior as a result of strategic shift in

---

[6] For example, if the New York AI Meetup hosted ten total events in 2010, and a member that entered the network in 2010 attended four of those events, we simulate that member as having a 40% probability of attending any New York AI Meetup event throughout the entire time period.

[7] "Year" is defined as 365 days after an individual's first event attendance. We simulate no more future attendances to events of a group that subsequently becomes inactive. We do this because we consider any exploration of new groups (and therefore, meeting new acquaintances) as a form of rewiring.

preferences or curiosity-driven purposive exploration of the network. However, the change in event attendance behavior itself is enough to realize potential knowledge spillover within different parts of the knowledge network and to enable innovative activity.

**Figure 6:** Modularities under simulations of differentiated preferences but no rewiring. The difference between observed and expected modularities can be attributed to members' changing preferences over time.

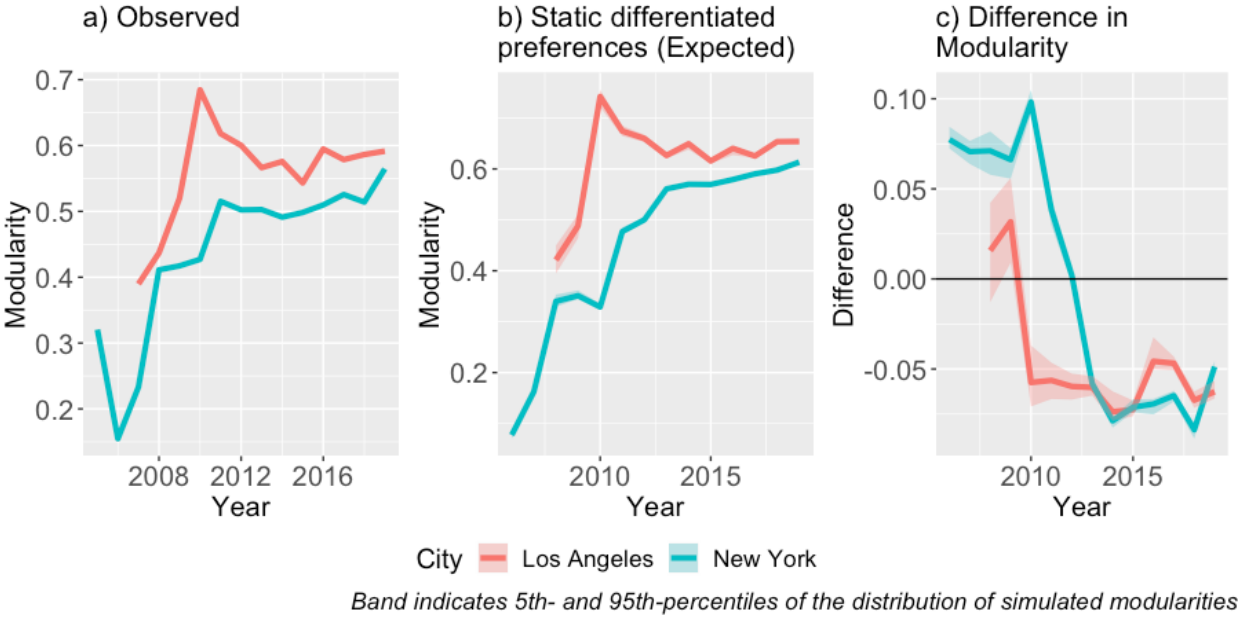

As the knowledge economy matures, heterogeneous preferences among a growing pool of members facilitate the formation of social boundaries of modular communities, but rewiring tends to integrate the network by creating bridge ties across modular communities. At the micro-level, as individuals select which events to pursue and which to forgo, they are individually exposed to more diverse interests through other individuals they encounter at their events. At the macro-level, this type of selection leads to more integrated knowledge clusters as rewiring and exploration provide bridges to new knowledge clusters that emerge as the network expands across economic institutions and organizations of the community ecology of a knowledge economy.

*Conclusion*

This paper introduces novel methods to quantify specialized knowledge and network rewiring using a unique data set from the professional networking platform Meetup.com. We measured specialized knowledge using individuals' self-reported interests, the interest tags associated with the groups they join, and their network position relative to others. The paper also demonstrates methods to quantify novelty that an individual may experience in a dynamic network, and simulated counterfactuals are used to quantify the impact of individual exploration on network structure.

As network analysis is increasingly used to visualize and analyze data, these methods will allow researchers to quantify fundamental ideas about knowledge spillovers and network rewiring. Application of these methods to networks of technologists in two metropolitan regions uncovered knowledge spillover and network rewiring of participant technologists and entrepreneurs. These new sources of data and methods enable us to see with greater clarity the network morphology of the division of knowledge in a technology economy as it undergoes both increasing specialization among those already present and also an influx of new people with different knowledge and interests.

*Appendix A:* Specialization score

Suppose member $i$ has $m_i$ interest terms, represented by the set $\boldsymbol{\tau}_i = \{\tau\}^{m_i}$ where $\tau \in T$ and $T$ is the set of all possible interest terms. The network $G(V,E)$ containing member $i$ as a node (i.e. $i \in V$) has a corresponding distribution of interests that reflect the popularity of each interest among the population of its nodes. This interest distribution can be thought of as a probability distribution with a mass function $f_G(\tau)$ that gives the "share" that every interest topic $\tau$ has on the overall population of members in the network. For example, in a population of 10 people, an interest that 3 of the people have has a "share" of 0.3. Let

$$\delta_i(\tau) = \begin{cases} 1 & \tau \in \boldsymbol{\tau}_i \\ 0 & otherwise \end{cases} \quad (A1)$$

Then formally,

$$f_G(\tau) = \frac{\sum_{i \in V} \delta_i(\tau)}{\sum_{i \in V} m_i} \quad (A2)$$

For every node $i$, its ego network consists of the focal node $i$ as well as all of its neighbors. Let $G_i$ be the set of node $i$ and all its neighbors, then we can define node $i$'s ego interest probability mass function as

$$f_{G_i}(\tau) = \frac{\sum_{j \in G_i} \delta_j(\tau)}{\sum_{j \in G_i} m_j} \quad (A3)$$

We define a member's specialization score as the difference in the member's ego interest distribution and the interest distribution of the entire network.

$$SpecializationScore_i = \sum_{\tau \in T} |f_G(\tau) - f_{G_i}(\tau)| \quad (A4)$$

The choice of distance metric is flexible. The use of absolute difference in this specification ensures that the specialization score is bounded between 0 and 2 since each probability mass function sums to 1.

Figure A1 shows a visualization of a sample of the 2019 New York member network where nodes are colored by the quartile of their specialization scores. We see that members with the highest specialization scores tend to be positioned in clusters near the periphery of the network while members with the lowest specialization scores tend to be those who are either connected to a large number of other members or are positioned outside of the most clearly defined clusters.

**Figure A1:** Visualization of specialization scores in the 2019 New York members networks. Note that nodes with the highest specialization scores tend to be positioned in more well-defined clusters.

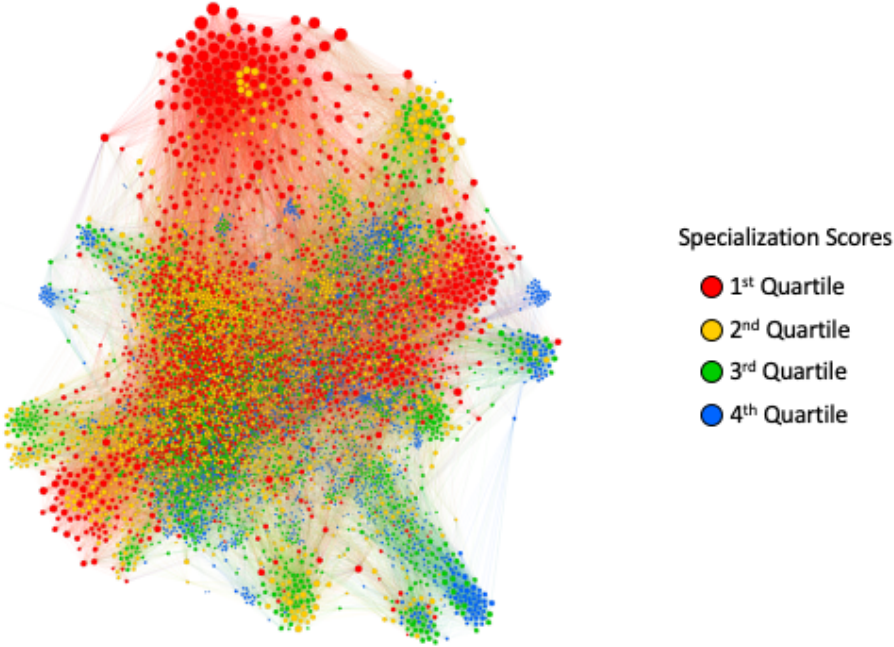

*Appendix B:* Modularity

Modularity is defined for a given partition on a given network and is the fraction of edges that fall within the given partitions minus the expected fraction if edges were distributed at random. Formally, the definition of network modularity on a network $G(V, E)$, where $V$ is the set of all nodes and $E$ is the set of all edges, and partition $C$ is given by Equation B1.

$$Q(G(V,E), C) = \frac{1}{2|E|} \sum_{v,w \in V} \left[ A_{vw} - \frac{k_v k_w}{2|E|} \right] \delta(c_v, c_w) \tag{B1}$$

$A$ is the adjacency matrix, and $k_v$ is the degree of node $v$ and $\delta(c_v, c_w)$ is an indicator function that is equal to 1 when nodes $v$ and $w$ fall into the same partition under $C$, and 0 otherwise.

This definition of modularity implicitly presents a way of detecting communities in a given network $G$ if we maximize over the set of all possible partitions on the network, $\mathbb{C}_G$. Hence, if we can solve the optimization given by Equation B2, we can find optimal communities in terms of this modularity measure; additionally, the value of the optimization gives a network-level statistic quantifying the network structure on the network $G$.

$$Q_{max}(G(V,E), C) = \max_{C \in \mathbb{C}_G} Q(G(V,E), C) \tag{B2}$$

This optimization, however, is equivalent to a linear integer program that is NP-hard. Various algorithms have been proposed to approximate the solution. We investigated several of these proposed algorithms. The Louvain algorithm consistently produces the highest values of network modularity among all the tried algorithms, and the Louvain algorithm is the one we use for the main modularity analysis.